# Technique detection software for Sparse Matrices

**Muhammad Taimoor Khan**
**Pakistan Institute of Engineering and Applied Sciences (PIEAS), Islamabad, Pakistan**
taimoor.muhammad@gmail.com

**Dr. Anila Usman**
**Pakistan Institute of Engineering and Applied Sciences (PIEAS), Islamabad, Pakistan**
Anila@pieas.edu.pk

**ABSTRACT.** Sparse storage formats are techniques for storing and processing the sparse matrix data efficiently. The performance of these storage formats depend upon the distribution of non-zeros, within the matrix in different dimensions. In order to have better results we need a technique that suits best the organization of data in a particular matrix. So the decision of selecting a better technique is the main step towards improving the system's results otherwise the efficiency can be decreased. The purpose of this research is to help identify the best storage format in case of reduced storage size and high processing efficiency for a sparse matrix.
**KEYWORDS.** Sparse matrices, sparse storage formats, sparse matrix vector multiplication

## 1 Introduction

Data compression is the main issue these days since data is being computerized and the size of data keeps on increasing. So various techniques have been devised to reduce the storage size of data in order to move and process it efficiently. Sparse storage formats steps to reduce the storage size of the data in a way not to lose the important information.

Sparse matrices are obtained from large linear systems that mostly consist of zeros. When these matrices are stored and processed as dense matrices they consume much more resources (memory and processing time). It gives birth to the idea of using some special techniques named as sparse storage formats to deal with such data. These storage formats are developed by different scientists and engineers after experimenting on the data deducted form large circuits, models, chemical reactions, linear problems, network traffic etc.

All of these storage formats followed the same theme of excluding the unnecessary zeros from the matrix data in a different way. Some additional information is also added to the matrix for preserving the original shape of the matrix. The performance of these techniques varies with the distribution of non-zeros within matrix. "No storage technique is efficient for all the sparse matrices; however, the selection of a suitable one gives better results" [DDR00].

Towards the goal of improving the efficiency of sparse matrices operations, a software has been implemented that suggests to the user which storage format suits best a particular matrix data, after it analyzes the matrix data. It is a step towards improving the results of an end user of sparse storage formats and encouraging more towards the safe and successful use of sparse storage formats.

In section 1 sparse storage formats are introduced and their importance is discussed. Section 2 explains sparse storage formats briefly. Selection of the most appropriate storage formats is described in section 3. Section 4 gives sample results of the software. The conclusions and future recommendations are in section 5 while section 6 consists of references.

## 2 Sparse Storage Formats

Those sparse storage formats in which the non-zeros are accessed from their location in the original matrix are called point entry storage formats. While, those storage formats in which the non-zeros are accessed through their blocks are called block entry storage formats. The following storage formats have been implemented and are considered in technique detection software. They are:

- Coordinate storage format [DDR00]
- Compressed column storage format [DDR00]
- Compressed row storage format [DDR00]
- Compressed diagonal storage format [DDR00]
- Jagged diagonal storage format [EM03]
- Transpose jagged diagonal storage format [EM03]

In order to have a better idea of the working of sparse storage formats, compressed row storage format is discussed. It also gives ratio of compression by comparing the matrix data in the compressed row with the dense matrix.

In compressed row storage format the non-zeros are stored along with their column indices and row pointer which points to the first nonzero of each row. The conversion of a matrix data from dense to compressed row storage format is show in Figure 2.1.

$$A = \begin{bmatrix} 2 & 1 & 0 & 0 \\ 0 & 4 & 3 & 5 \\ 7 & 0 & 6 & 0 \\ 0 & 0 & 0 & 8 \end{bmatrix}$$

**Figure 2.1: Matrix data in Dense and Compressed Row storage format**

| Value | 2 | 1 | 4 | 3 | 5 | 7 | 6 | 8 |
|-------|---|---|---|---|---|---|---|---|
| Col index | 1 | 2 | 2 | 3 | 4 | 1 | 3 | 4 |
| Row ptr | 1 | 3 | 6 | 8 | 9 | | | |

The purpose of sparse storage format is to reduce the storage size and increase the processing efficiency. The size comparison of different storage formats, their computations in matrix-vector multiplication is given below.

Let's consider

Matrix A = M x N
Vector =N x 1

NNZ is the total number of non-zeros in the matrix, NZD is the number of nonzero diagonals, JD is the jagged diagonal vector, and TJD is the transpose jagged diagonal vector while NPR is the number of non-zeros per row. The number of non-zeros per row is assumed to be equal for all rows. The table 2.1 compares the number of computations of a dense matrix to that of the coordinate storage format, which is the simplest of all.

**Table 2.1: Computations comparison in matrix-vector multiplication**

| Storage format | Computations |
|----------------|--------------|
| Dense | M x (2N -1) |
| COO | NNZ + M(Npr-1) |

The table 2.2 gives size comparison of dense matrix to that of the implemented sparse storage formats. The size comparison of sparse storage formats depends upon the size of the constants like NZD, JD, TJD etc. However, in case of sparse matrices they all perform better than the dense matrix.

**Table 2.2: Size comparison of different storage formats**

| Storage format | Size |
|---|---|
| Dense Matrix | M x N |
| COO | 3 x NNZ |
| CSR | 2 x NNZ + M + 1 |
| CSC | 2 x NNZ + N + 1 |
| CDS | if (M <= N)          NZD + (NZD x M)<br>Else                       NZD + (NZD x N) |
| JDS | ( 2 x NNZ ) + (JD +1) + M |
| TJDS | ( 2 x NNZ ) + (TJD +1) |

## 3 Selection of appropriate storage format for a particular matrix

The above mentioned sparse storage formats were implemented and tested with matrices of all possible data distributions. The size of the selected matrices is compared in all implemented storage formats. Matrix-vector multiplication was applied on them to compare their performance. The results obtained leads us to the conclusion that, for each data distribution the technique that performs well, can be used for the matrices that have data distribution similar to it. A set of rules has been devised to identify the category of the matrix data distribution and then to suggest the most appropriate storage format. The categories in which we can distribute the matrix data are row wise, column wise and diagonal density and randomness. A rule to find the randomness in the matrix data is shown in Table 3.1. A matrix data is ideally random if all the non-zeros are at equal distance from each other in their respective rows and columns. Those non-zeros that are ideally random are counted to calculate, its percentage by comparing with the total number of non-zeros within the matrix. For a matrix of nonzero density of 20 percent, the ideal random spacing is 80 percent between non-zeros.

**Table 3.1: Percentage of ideally random non-zeros in the matrix**

| Randomness Level | Percent density in the ideal case |
|---|---|
| Very High | 75 percent or above |
| High | 60-75 percent |
| Medium | 50-60 percent |
| Low | 40-60 percent |
| Very Low | 40 percent or below |

The rule for suggesting a row or column as dense row or dense column is same. If in a row or column there are more non-zeros more than half of its length then it is considered as dense row or column. The numbers of non-zeros in dense rows or dense column are counted. It gives us the percentage of non-zeros in dense rows or columns. In order to consider a matrix as dense row or dense column matrix depends upon its percentage which specifies a category to the matrix as listed in Table 3.2.

**Table 3.2: Percentage of non-zeros in dense rows/columns in the matrix**

| Density Level | Percent density |
|---|---|
| Very High | 50 percent and above |
| High | 35 percent to 50 percent |
| Medium | 25 percent to 35 percent |
| Low | 15 percent to 25 percent |
| Very Low | Below 15 percent |

Shape of the matrix or its dimensions also affects the efficiency of a storage formats. There are three possibilities for the shape of a matrix. It can be square, horizontally rectangular or vertically rectangular. The affect of shape is also included in consideration while selecting a storage format.

Since in a matrix there are rows+cols-1 possible diagonals. So if there is low density of nonzero diagonals, then it's obvious that the density of non-zeros in the nonzero diagonals will be high. We can also calculate the diagonal density from the number of non-zeros within nonzero diagonals. But in case of diagonal density we need less number of nonzero diagonals. The reason is that if we have even 90% of non-zeros in dense diagonals but if the remaining 10% are distributed widely over nonzero diagonals the efficiency will decrease. Table 3.3 shows a list of nonzero diagonals of the matrix against their density levels.

**Table 3.3: Percentage of nonzero diagonals in the matrix**

| Density Level | Percent density |
|---|---|
| Very High | 15 percent or below nonzero diagonals |
| High | 15 percent to 30 percent nonzero diagonals |
| Medium | 30 – 45 percent nonzero diagonals |
| Low | 45-60 percent nonzero diagonals |
| Very Low | Above 60 percent nonzero diagonals |

When the matrix is analyzed the storage formats is suggested to it, according to its density in each of the above discussed categories. The set of rules on the basis of which a sparse storage format has been selected after analyzing the matrix data for the objective of high processing efficiency is shown in Table 3.4:

**Table 3.4: Suitable storage formats for data distributions
to obtain high processing efficiency**

| Density Level | Category | Shape | Technique |
|---|---|---|---|
| Very High | Diagonal Density | Square/ Vertically rectangular/ Horizontally rectangular | DIA |
| Very High | Dense Rows | Square/ Vertically rectangular | CSR |
| Very High | Dense Columns | Square/ Vertically rectangular | JDS |
| Very High | Randomness | Square/ Vertically rectangular/ Horizontally rectangular | TJDS |
| Very High | Dense Rows | Horizontally rectangular | TJDS |
| Very High | Dense Columns | Horizontally rectangular | CSC |

If the objective is to reduce the storage size of the matrix storage formats are listed against different data distributions. Coordinate storage format is the only among the implemented storage formats whose storage size is not affected by the data distribution within the matrix. Obviously the number of non-zeros affects all the storage formats. The set of rules that help to suggest the most appropriate storage format in case of reduced storage size is shown in Table 3.5:

**Table 3.5: Suitable storage formats for different data distributions to reduce the storage size**

| Density level | Category | Shape | Technique |
|---|---|---|---|
| Very High | Diagonal Density | Square/ Vertically rectangular/ Horizontally rectangular | DIA |
| Very High | Randomness | Square/ Vertically rectangular/ Horizontally rectangular | TJDS |
| Very High | Dense Rows | Square/ Horizontally rectangular | CSR |
| Very High | Dense Rows | Vertically rectangular | CSC |
| Very High | Dense Columns | Square/ Vertically rectangular | CSC |
| Very High | Dense Columns | Horizontally rectangular | JDS |

## 4 Sample results

This section describes sample results when "Technique detection software for sparse matrices" was provided with input matrices (Mbeause and lrand). In the first portion the matrix data is shown. Then the analysis on the matrix data is displayed. While at the end the most appropriate storage format is suggested to achieve the goals of reduced size and high efficiency. The storage format is suggested among the implemented ones. A sample output of the software is provided in Figure 4.1 which shows the matrix data being parsed and suggested the best storage format to achieve high processing efficiency as well as a sparse storage format to reduce the storage size of the matrix to minimum.

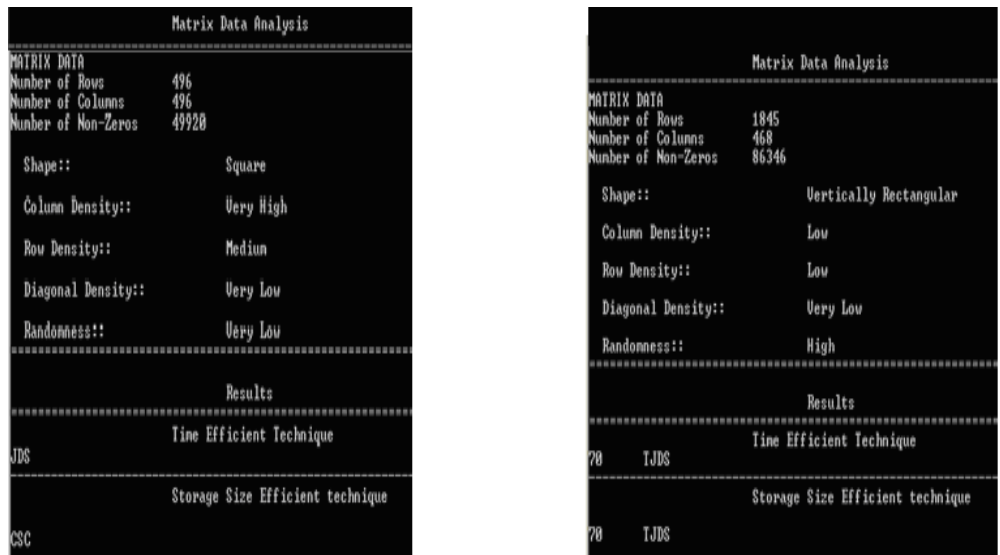

**Figure 4.1: Sample results of technique detection software
for sparse matrices**

## 5 Conclusion and Future Recommendations

The performance of sparse storage format is affected by the number of non-zeros and the distribution of non-zeros within the matrix. Selection of the most appropriate technique improves the processing efficiency and reduces storage size. While in some cases we will have to select one at the cost of the other depending upon the requirements of the user.

More sparse storage formats can be implemented and tested on a set of matrices having almost all possible data distributions to obtain their areas of brilliance. Rules can added for those storage formats in the software, in order to provide the end user with more number of options and to have a more specific technique for a particular data distribution.